\documentclass[12pt]{article} 
\usepackage{graphics,cite,amssymb,epsfig,float,psfrag} 
\usepackage[usenames,dvips]{color} 
\usepackage{rotating} 
 
\begin{document} 
\newcommand{\lsim}{\raisebox{-0.13cm}{~\shortstack{$<$ \\[-0.07cm] $\sim$}}~} 
\newcommand{\gsim}{\raisebox{-0.13cm}{~\shortstack{$>$ \\[-0.07cm] $\sim$}}~} 
\newcommand{\ra}{\rightarrow} 
\newcommand{\lra}{\longrightarrow} 
\newcommand{\ee}{e^+e^-} 
\newcommand{\gam}{\gamma \gamma} 
\newcommand{\nn}{\noindent} 
\newcommand{\non}{\nonumber} 
\newcommand{\beq}{\begin{eqnarray}} 
\newcommand{\eeq}{\end{eqnarray}} 
\newcommand{\s}{\smallskip} 
\def\NPB{Nucl. Phys. B} 
\def\PLB{Phys. Lett. B} 
\def\PRL{Phys. Rev. Lett.} 
\def\PRD{Phys. Rev. D} 
\def\ZPC{Z. Phys. C} 
\baselineskip=16pt 

\rightline{LPT--07-15} 
\rightline{LPSC--07-26} 
 
\vspace{0.7cm} 
 
\begin{center}

{\Large {\bf Probing RS scenarios of flavour at LHC via leptonic channels}} 
 
\vspace{0.7cm} 
 
{\large Fabienne Ledroit $^1$, Gr\'egory Moreau $^2$, Julien Morel $^1$} 
 
\vspace{0.7cm} 
 
1: LPSC, Universit\'e Joseph Fourier Grenoble 1, CNRS/IN2P3, 
Institut National Polytechnique de Grenoble, Grenoble, France \\ 
2: Laboratoire de Physique Th\'eorique, CNRS and Universit\'e Paris--Sud, \\ 
B\^at. 210, 91405 Orsay, France

\end{center} 
 
\vspace{0.5cm} 
 
\begin{abstract}  
We study a purely leptonic signature of the Randall--Sundrum scenario 
with Standard Model fields in the bulk at LHC:  
the contribution from the exchange of Kaluza--Klein (KK) 
excitations of gauge bosons to the clear Drell--Yan reaction.
We show that this contribution is 
detectable (even with the low luminosities of the LHC initial regime) 
for KK masses around the TeV
scale and for sufficiently 
large lepton couplings to KK gauge bosons. 
Such large couplings can be compatible with ElectroWeak precision 
data on the $Z \bar f f$ coupling   
in the framework of the custodial $O(3)$ symmetry recently proposed, for 
specific configurations of lepton localizations (along the extra dimension).  
These configurations can simultaneously reproduce the correct lepton masses,  
while generating acceptably small Flavour Changing Neutral Current (FCNC) effects.    
This LHC phenomenological analysis
is realistic in the sense that it is based on fermion localizations  
which reproduce all the quark/lepton masses plus 
mixing angles and respect FCNC constraints in both the hadron and lepton sectors. 
\end{abstract} 
 
\newpage

\section{Introduction}  
\label{intro} 
 
Among the recent extra--dimensional effective scenarios, the one    
proposed by Randall and Sundrum (RS) \cite{RS}, based on an additional warped dimension,   
seems quite attractive. The RS scenario provides 
a favorable framework for alternative models of ElectroWeak (EW) symmetry breaking, 
like the Higgsless \cite{Higgsless}, gauge--Higgs unification \cite{gaugehiggs} 
or composite Higgs \cite{MHCM} models.  
From a more generic point of view, the RS scenario can address 
the gauge hierarchy problem without introducing any new energy scale  
in the fundamental theory. Moreover, the variant of the original RS model,  
with Standard Model (SM) fermions and bosons propagating in the bulk, 
allows for the unification of gauge coupling constants at a high 
energy Grand Unification scale \cite{UNI-RS} and provides  
viable candidates of Kaluza--Klein (KK) type for the dark matter of 
the universe \cite{LZP2}.  
\\ In this version of the RS model with bulk matter, 
a purely geometrical origin arises naturally   
for the large mass hierarchies prevailing among SM fermions 
\cite{RSloc,HSquark,GNeubertA}. The principle is that if the various SM fermions 
are displaced along the extra dimension, 
their different wave function overlaps with the Higgs boson (which remains 
confined on the so--called TeV--brane for its mass to be protected) generate  
hierarchical patterns among the effective 4--dimensional Yukawa couplings. 
With such a geometrical approach, the quark masses and CKM mixing angles can be  
accommodated \cite{HSquark}, as well as the lepton masses and MNS  
mixing angles in both cases where neutrinos have masses of type Majorana  
\cite{HSHHLL} or Dirac \cite{HSpre,RSflav} 
\footnote{There are other higher--dimensional mechanisms  
\cite{GNeubertA}, in the context of warped extra--dimensions, applying specifically to neutrinos   
and explaining their relative lightness.}.\s

In the framework of the RS model with bulk fields, if the gauge hierarchy problem 
is to be solved, the mass of the first KK excitation of SM gauge bosons must be  
of order of the TeV scale.  
Hence, KK excitations of gauge bosons are expected to be produced significantly 
at the forthcoming Large Hadron Collider (LHC), which provides a center-of-mass  
energy of $14$ TeV, for KK gauge boson couplings to light quarks of the same  
order as the SM gauge couplings 
\footnote{In the RS context, light KK excitations  
of quarks \cite{Servant} as well as KK gravitons \cite{Kaplan,Dav} 
can also be produced significantly at LHC (or SLHC).}.\s

In the present work, we develop a test of KK excitation effects at LHC, in the RS  
scenario with bulk fields generating the SM fermion masses: we study the direct 
contributions of KK excitations of the photon and of the $Z$ boson to the SM Drell--Yan process, namely 
$p p \to \gamma^{(n)} / Z^{(n)} \to \ell^+ \ell^-$, $n$ being the KK--level.  
The motivation for considering this 
process is that the neutral KK excitations can be produced as resonances, tending to increase  
considerably the total amplitude. Moreover, the di--lepton final state constitutes a  
particularly clean signature in an hadronic collider environment. 
\\ In the framework of the RS model with bulk matter, the high energy collider phenomenology  
and flavour physics are interestingly connected: the effective 4--dimensional  
couplings between KK gauge boson modes and SM fermions depend on fermion localizations  
along the extra--dimension which are fixed (non uniquely) by fermion masses. 
In the present study for the LHC,  
this connection between collider and flavour physics will be taken into account 
as we will consider some fermion location configurations which reproduce all the 
quark/lepton masses and mixing angles, and, satisfy Flavour Changing Neutral Current~(FCNC) 
constraints for masses of the first KK gauge bosons around the TeV scale  
(see Ref.~\cite{HSquark,RSflav,AgasheTH} for general discussions on these FCNC effects 
and Ref.~\cite{AgasheEXP,BurdmanEXP} for experimental status). 
This is in contrast with the preliminary study \cite{RizzoUniv} 
on the reaction $p p \to \gamma^{(n)} / Z^{(n)} \to \ell^+ \ell^-$  
in the RS model, which was performed under 
the assumption of universal fermion locations (in order to totally avoid FCNC effects)  
so that SM fermion mass hierarchies were not able to be generated.\s

Usually, the production of heaviest SM fermions (typically localized towards the TeV--brane to have a   
large overlap with the Higgs boson) are considered to be favored due to their larger couplings to 
KK gauge bosons (also located near the TeV--brane). This has motivated recently the study, in the RS model, of     
the top quark pair production at LHC (through direct KK gluon production)
\cite{Agashe:2006hk,Wang} and ILC (via virtual $\gamma^{(n)} / Z^{(n)}$ exchanges) \cite{Sher}. 
Nevertheless, as will be discussed, if the left--handed charged leptons are localized closely to the TeV--brane 
whereas the right--handed ones are rather close to the Planck--brane, the lepton masses can still be small enough 
and compatible with significant couplings between left--handed charged leptons and KK gauge bosons.    
Such large KK couplings of leptons could be in agreement with the constraints from 
the EW precision data on $Z \ell \bar \ell$ vertex 
if one assumes a custodial symmetry \cite{Agashe:2003zs,Djouadi:2006rk} and more precisely an  
$O(3)$ symmetry \cite{O3cust,gaugehiggs}. This $O(3)$ symmetry will also allow to generate 
the heavy top mass, and simultaneously, protect the $Z b \bar b$ coupling as well as $\Delta \rho$  
against too large corrections from KK state exchanges 
(the elimination of this tension was the original motivation  
for introducing the $O(3)$ symmetry \cite{O3cust}).   
Hence, the leptonic signature which is studied here is characteristic of the phenomenology of the 
RS scenario with a custodial $O(3)$ symmetry.\s

The paper is organized as follows. In next section, the
theoretical context is described, whereas in Section~\ref{PhenoBounds}, the relevant 
phenomenological constraints are discussed. The search at LHC is
studied in Section~\ref{LHC}, and conclusions are drawn in Section~\ref{conclu}.\s

\section{Theoretical framework}  
\label{fundamental}

We begin by discussing the values of fundamental parameters in the RS model.  
While on the Planck--brane the effective gravity   
scale is equal to the (reduced) Planck mass: $M_{Pl}=2.44\   
10^{18}$ GeV, on the TeV--brane the gravity scale,   
$M_{\star }=w\ M_{Pl}$,  
is suppressed by the exponential `warp' factor $w=e^{-\pi kR_{c}}$,  
where $1/k$ is the curvature radius of Anti--de--Sitter space and $R_c$  
the compactification radius.   
For a small extra dimension   
$R_{c}\simeq 11/k$ ($k$ is taken close to $M_{Pl}$),   
one finds $w\sim 10^{-15}$ so that $M_{\star}={\cal O}(1)$ TeV,   
thus solving the gauge hierarchy problem.   
Solving the gauge hierarchy problem forces $M_{KK}$ 
(the mass of the first KK excitation of SM gauge bosons: $M_{KK} 
=M_{\gamma^{(1)}} \simeq M_{Z^{(1)}}$) 
to be of order of the TeV scale. Indeed, one has 
$M_{KK}=2.45 k M_{\star}/M_{Pl} \lesssim M_{\star} = {\cal O}(1)$ TeV since 
the theoretical consistency bound on the 5--dimensional curvature scalar 
leads to $k<0.105 M_{Pl}$. 
More precisely, the maximal value of $M_{KK}$ is fixed by this theoretical 
consistency bound and the $kR_{c}$ value. One could consider a maximal 
value of $M_{KK} \simeq 10$ TeV which corresponds to $kR_{c}=10.11$. 
Since we are interested in the search for KK state effects at   
LHC, $M_{KK}$ will be taken instead of $k$ as the free parameter, which is equivalent.\s

Concerning the mass values for the SM fermions,  
they are dictated by their wave function location. 
In order to control these locations, the 5--dimensional fermion fields 
$\Psi_{i}$ (the generation index $i=\{1,2,3\}$) are usually coupled to 
distinct masses $m_{i}$ in the fundamental theory. If $m_{i}= {\rm 
sign}(y) c_{i} k$, where $y$ parameterizes the fifth dimension and 
$c_{i}$ are dimensionless parameters, the fields decompose as 
$\Psi _{i}(x^{\mu },y)= \sum_{n=0}^{\infty }\psi_{i}^{(n)}(x^{\mu }) 
f_{n}^{i}(y)$, where $n$ labels the tower of KK excitations and 
$f_{0}^{i}(y)=e^{(2-c_{i})k|y|} / N_{0}^{i}$ ($N_{0}^{i}$ being just 
a normalization factor). Hence, as $c_i$ increases, the wave function 
$f_{0}^{i}(y)$ tends to approach the Planck--brane at $y=0$.\s

We finish this section by recalling how the locations of fermions fix their 
effective 4--dimensional couplings to KK gauge bosons.  
The neutral current action of the effective 4--dimensional coupling,   
between SM fermions $\psi_{i}^{(0)}(x^{\mu})$   
and KK excitations of any neutral gauge boson $A_{\mu}^{(n)}(x^{\mu})$,   
reads in the interaction basis as,   
\begin{equation}   
S_{\rm NC}= g_L^{SM} \int d^4x \sum_{n=1}^{\infty}   
\bar \psi_{Li}^{(0)} \ \gamma^{\mu} \ {\cal C}_{Lij}^{(n)} \ \psi_{Lj}^{(0)} \ A_{\mu}^{(n)}   
\ + \ \{ L \leftrightarrow R \},   
\label{Sweak}   
\end{equation}   
where $g_{L/R}^{SM}$ is the relevant SM gauge coupling constant and   
${\cal C}_{Lij}^{(n)}$ the $3 \times 3$ diagonal matrix  
$diag(C_0^{(n)}(c_1),C_0^{(n)}(c_2),C_0^{(n)}(c_3))$.  
These factors $C_0^{(n)}(c_i)$ quantify the wave function overlap (along the extra  
dimension) between the  
localized KK excitation of gauge boson $A_{\mu}^{(n)}$ and the localized SM  
fermions $\psi_{i}^{(0)}$.  
In case of the RS model, the expression for coefficient $C_0^{(n)}(c_i)$ is given e.g.  
by the coefficient $C_{00n}^{f_i \bar f_i A}$ defined in Ref.~\cite{RizzoUniv}.   
\\ The action in Eq.(\ref{Sweak}) can be rewritten in the mass basis (indicated   
by the prime):   
\begin{equation}   
S_{\rm NC}= g_L^{SM} \int d^4x \sum_{n=1}^{\infty}   
\bar \psi_{L\alpha}^{(0) \ \prime} \ \gamma^{\mu} \ V_{L\alpha\beta}^{(n)}   
\ \psi_{L\beta}^{(0) \ \prime} \ A_{\mu}^{(n)}   
\ + \ \{ L \leftrightarrow R \},   
\label{Sphysical}   
\end{equation}   
\begin{equation}   
V_{L}^{(n)}=U_L^\dagger \ {\cal C}_{L}^{(n)} \ U_L,   
\label{Vmatrix}   
\end{equation}   
$U_L$ being the unitary matrix of basis transformation for left--handed fermions   
and $\alpha,\beta$ being flavour indices. 
One can see that the non--universality of the   
effective coupling constants $g_{L/R}^{SM} \times C_0^{(n)}(c_i)$ between KK modes   
of the gauge fields and the three SM fermion families (which have different   
locations along $y$), in the interaction basis, induces non vanishing off--diagonal   
elements for matrix $V_{L/R}^{(n)}$, in the mass basis, giving rise   
to Flavour Changing (FC) couplings.\s

\section{Phenomenological constraints}   
\label{PhenoBounds}

\noindent $\bullet$ {\bf Fermion masses:}  
In this paper, for the purpose of illustration, 
three characteristic examples of complete sets for the $c_i$ parameter values
are considered: the sets A, B and C presented in the Appendix.   
\\ The three fermion localization configurations, corresponding to sets A, B and C,   
have been shown in \cite{RSflav} to reproduce all the present data on   
quark/lepton masses and mixing angles (in case of Dirac neutrino masses induced by the presence  
of three right--handed neutrinos), through the geometrical mechanism \cite{RSloc} described in Section~\ref{intro}. 
The effective quark/lepton mass matrices, generated via this mechanism, depend   
on the $c_i$ and  
the RS parameter product $kR_{c}$, which was fixed in \cite{RSflav} to the same amount as here.  
\\ In particular, for these three sets,  
the unusually low $c_i^L$ values ($c_i^L<0.5$) for left--handed charged leptons are compensated by 
some large $c_i^\ell$ values for right--handed ones so that the correct electron, muon and tau 
masses can be generated.\s

\vskip .5cm  
  
\noindent $\bullet$ {\bf FCNC effects:} 
The indirect phenomenological constraints on $M_{KK}$  
holding in the RS model with bulk matter must be considered. 
The experimental limits on FCNC processes translate into a lower bound on $M_{KK}$.  
Indeed, within the context of the RS scenario creating fermion masses, FCNC processes are induced at   
tree level by exchanges of KK excitations of neutral gauge bosons. This is 
rendered possible by the fact that these KK states   
possess FC couplings to fermions ({\it c.f.} Eq.(\ref{Sphysical})).  
This is necessary as the mass hierarchies and mixings of SM fermions  
require flavour and nature dependent locations for quarks/leptons, 
or equivalently (as described in previous section), different $c_i$ parameter values.  
\\ The FC couplings between KK gauge bosons and SM fermions are significantly suppressed for   
$c_i$ values corresponding to certain configurations of fermion localizations \cite{RSflav} 
(see also \cite{RSloc,AgasheTH}). For   
these localization configurations, experimental limits on KK--induced FCNC effects are   
satisfied even for rather low KK masses. Sets A, B, C of $c_i$ values given in the Appendix   
correspond to such configurations: for these three sets of $c_i$ values,   
it was shown in \cite{RSflav} that FCNC reactions in both the hadron and lepton   
sector (like $b \to s \gamma$, $B^0-\bar B^0$, $\mu^- \to e^- e^+ e^-$,   
$K \to \mu^+ \mu^-$,\dots) respect the experimental limits if $M_{KK} \gtrsim 1$ TeV.\s

\vskip .5cm

\noindent $\bullet$ {\bf EW measurements:}  
Secondly, the mixing between the EW gauge bosons and their KK modes induces modifications   
of the boson masses/couplings, and thus deviations to EW precision observables
\footnote{See \cite{Falkowski} for the discussion of EW observables in a general warp background.}. 
Hence, the fit of EW precision data imposes  
the typical bound $M_{KK} 
\stackrel{>}{\sim} 10$ TeV \cite{RizzoUniv,Burdman:2002gr}. Thus we first consider   
the scenario with the EW gauge symmetry enhanced to $SU(2)_{L} \times  
SU(2)_{R} \times U(1)_{X}$ \cite{Agashe:2003zs} 
\footnote{Another kind of scenario was suggested in the literature in order to 
relax the EW bound on $M_{KK}$ down to a few TeV:  
the scenario with brane localized kinetic terms for fermions \cite{BraneF}   
or gauge bosons \cite{BraneB} (see \cite{EWBa}   
for gauge boson kinetic terms and \cite{EWF} for fermion ones).} 
leading to reasonable  
fit of the oblique S,T parameters for $M_{KK} \stackrel{>}{\sim} 3$ TeV  
and the $c_i^{f_R}$ (for right--handed SM fermions) configurations considered in our A, B, C sets,
namely $c_{i}^{d,\ell,\nu}>0.5$, $c_{1,2}^{u}>0.5$, $c_{3}^{u}<0.5$ 
($i=1,\dots,3$ being the generation index).  
In the three sets, 
the low $c_{3}^{u}$ and $c_{3}^{Q}$ values (pushing typically the $t_{L/R}$,$b_L$ 
towards the TeV--brane), needed to generate the large top mass, give rise to significant $b_L$ 
couplings to KK gauge bosons. So in order to force the deviations  
(from both the mixing with KK gauge bosons and KK fermions)  
of the $Z \bar b_L b_L$ coupling to vanish for any $c_{3}^{Q}$ value,  
while still protecting the $\rho$ parameter  
against radiative corrections (by the already mentioned custodial $O(3)$ symmetry),  
the third family left--handed SM quark doublet $Q^3_L$ is embedded in  
a bidoublet $({\bf 2},{\bf 2})_{2/3}$ under the extended EW symmetry, as proposed  
in \cite{Agashe:2006at} and in contrast with \cite{Agashe:2003zs}.  
The two other $Q^{1,2}_L$ light quark doublets are  
also embedded in bidoublets $({\bf 2},{\bf 2})_{2/3}$.  
Then the $u^{i}_R$ quarks must belong to a representation corresponding to
$I_{3R}(u^{i}_R)=I_{3L}(u^{i}_R)=0$, which protects the $Z \bar u^{i}_R u^{i}_R$ vertex
against any KK contribution \cite{Agashe:2006at}.
As suggested recently in \cite{gaugehiggs}, 
the three families of left--handed SM lepton doublets $L^{i}_L$ are similarly embedded
into bidoublets $({\bf 2},{\bf 2})_0$. 
This guarantees that there are no modifications of the $Z \bar e_L e_L$, $Z \bar \mu_L \mu_L$ 
and $Z \bar \tau_L \tau_L$  
couplings, even for our chosen relatively low $c_{i}^{L}$ values  
that lead to a significant enhancement in the couplings 
between left--handed charged leptons and KK gauge bosons.
If light fermions are localized far from the TeV--brane, the S parameter is
positive as shown in \cite{gaugehiggs} (within the gauge--Higgs unification framework).
A precise analysis would be required for the case $c_{1,2}^{Q,L}<0.5$ 
(in the limit $c=0.5$ fermion couplings to KK gauge bosons vanish).
The set A has $c^L$ values much smaller than $0.5$ and should be excluded by EW 
constraints, but we just consider it in order to illustrate a strong coupling regime.\s

Let us describe more precisely the lepton charges/representations under the enhanced 
EW gauge group $SU(2)_{L} \times SU(2)_{R} \times U(1)_{X}$ (see \cite{Agashe:2006at} for the quark sector). 
The protection of the $Z \bar \ell^i_L \ell^i_L$ couplings requires the equality 
$I_{3R}(\ell^i_L)=I_{3L}(\ell^i_L)$ between the $SU(2)_R$ and $SU(2)_L$ isospin quantum numbers of the  
charged leptons.  
Hence, $Q_X(\ell^i_L)=0$ since the charge under $U(1)_{X}$ is related to the 
SM hypercharge $Y$ (given by $Q_{\rm em}-I_{3L}$) through: $Y=Q_X+I_{3R}$.                                    
Now, if the Yukawa term for charged leptons is issued from the minimal 
invariant operator with the form, 
\begin{equation} 
({\bf 2},{\bf 2})^H_0 \overline{({\bf 2},{\bf 2})_0} ({\bf 1},{\bf 3})_0 
\label{eq:min} 
\end{equation} 
where $({\bf 2},{\bf 2})^H_0$ represents the 
Higgs boson multiplet, then $\ell^{i}_R \in ({\bf 1},{\bf 3})_0 
\oplus ({\bf 3},{\bf 1})_0$ with $I_{3R}(\ell^i_R)=-1$. 
The $\ell^{i}_R$ representation could chosen differently at the price of 
generating the charged lepton masses by a non minimal operator, namely not as in Eq.(\ref{eq:min}) 
(an analog modification was proposed in \cite{Djouadi:2006rk,Agashe:2006at}
for $b_R$ 
in order to solve the forward--backward anomaly of the bottom quark).  
\\ For the neutrinos, one has $I_{3R}(\nu^i_L)=I_{3R}(\ell^i_L)$ and, similarly,   
the minimal operator for the Yukawa term  
(neutrino masses of Dirac type are considered along this paper) has the 
following invariant form,  
\begin{equation} 
({\bf 2},{\bf 2})^H_0 \overline{({\bf 2},{\bf 2})_0} ({\bf 1},{\bf 1})_0 
\ \ \mbox{or} \ \
({\bf 2},{\bf 2})^H_0 \overline{({\bf 2},{\bf 2})_0} ({\bf 1},{\bf 3})_0
\label{eq:minBIS} 
\end{equation}  
where $\nu^{i}_R \in ({\bf 1},{\bf 1})_0$ 
or $\nu^{i}_R \in ({\bf 1},{\bf 3})_0 \oplus ({\bf 3},{\bf 1})_0$, respectively, 
with $I_{3R}(\nu^i_R)=0$.\s

\section{LHC investigation}  
\label{LHC}

In the following, the A, B, C sets of $c_i$ parameters  
have been considered. The important connection is that these 
$c_i$ values, determining the SM fermion wave function profiles, fix    
the strength of couplings between SM fermions and KK gauge bosons which dictates 
the amplitude of KK effects at LHC.  
Indeed,  
the dependence of this strength (Eq.(\ref{Sphysical})) on the $c_i$ parameters 
enters (Eq.(\ref{Vmatrix})) via the ${\cal C}_{L/R}^{(n)}$ matrix  
as well as the $U_{L/R}$ matrices which diagonalize fermion mass matrices. 

Only the $M_{KK}\in[3,10]$ TeV range has been considered in order to 
simultaneously address the gauge hierarchy problem (see Section~\ref{fundamental}) and  
take into account 
the phenomenological constraints from FCNC processes as well as EW precision data  
(see Section~\ref{PhenoBounds}).\s 

In order to compute cross sections and to generate events, 
the $p p \to \gamma^{(n)} / Z^{(n)} \to \ell^+ \ell^-$ process
has been implemented as a user defined process in the {\tt PYTHIA} Monte Carlo
generator version~6.205~\cite{pythia}. 
Only the first three modes (i.e. up to the second KK excitation of the photon and of the $Z$ boson) 
were taken into account, as well as the interference between them. 
The contributions of $\gamma^{(n)}$, $Z^{(n)}$, with $n \geq 3$,
to the Drell--Yan cross section are not significant because the mass (fermion couplings) 
of $\gamma^{(n)}$, $Z^{(n)}$ increases (decreases) as the KK--level $n$ gets higher~\cite{RizzoUniv}.  
The second KK mass is already at $M_{\gamma^{(2)}}=(5.57/2.45) M_{KK}$, and
the third one is even higher.

The CTEQ5L\cite{cteq} Parton Density Functions (PDF) have been used.
Initial and final state radiation effects were included.\s   
 
\subsection{Cross sections and invariant mass distributions}   
\label{Cross}

\begin{figure}
\begin{center} 
\psfrag{Lumi}{}
\psfrag{XS}[br]{Cross section (fb)}
\psfrag{MZp}[r][b]{$M_{KK}$ (TeV)}
\psfig{figure=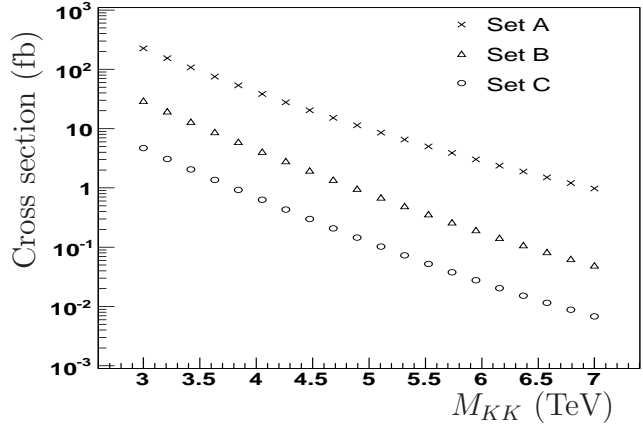,height=6cm,width=9cm} 
\end{center} 
\caption{ 
Cross section of the 
$p p \to \gamma^{(1,2)} / Z^{(1,2)} \to \ell^+ \ell^-$ process ($\ell=e$ only or $\mu$ only) at LHC as a function of
$M_{KK}$ for the three parameter sets A, B, C.}
\label{fig:xsec}
\end{figure}

The cross sections of the $pp \to \gamma^{(1,2)}/Z^{(1,2)} \to \ell^+\ell^-$ process alone 
(without the SM Drell--Yan contribution)
computed with {\tt PYTHIA} are shown as a function of $M_{KK}$
for the three parameter sets A, B and C  in Fig.~\ref{fig:xsec}.

The $c^L_i$ parameters considered here are almost universal in the family
space (namely for $i=1,2,3$) 
so that the wave function overlaps of left--handed leptons with KK
gauge bosons, and thus the effective leptonic couplings to KK gauge bosons, are
quasi identical. 
Furthermore, the $c^\ell_i$ are larger than $0.5$ and by consequence yield
almost universal KK gauge couplings to right--handed leptons. Indeed, for
$c \gg 0.5$, the ratio of KK over SM gauge coupling is fixed at $\sim -0.2$
since the KK gauge boson wave functions are quasi constant near the
Planck-brane. Therefore, the cross sections for the different lepton 
generations are practically equal, after having also taken
into account the dependence of
effective KK gauge couplings on lepton mixing angles (parameterizing 
the $U$ matrices of Eq.(\ref{Sphysical})-(\ref{Vmatrix})).
\newline On the other hand, one can see that the cross section gets 
higher when moving from set C to set B, and then to set A. 
The reason is that, the $c^{Q,L}_i$ values of set C are larger (this is not the case for the   
right--handed top quark, or more precisely $c_{3}^{u}$, but the top is not
involved in the studied reaction) than in set B and in turn larger than in set A, so that for  
this latter set the left--handed light fermions   
are localized closer to the TeV--brane, where are also located   
KK gauge bosons, leading to larger KK gauge couplings.
Concerning the other $c$ parameters, those are larger than $0.5$ leading to
almost universal KK gauge couplings, as already discussed.\s

\begin{figure} 
\begin{center} 
\psfrag{Lumi}{${\cal L}$ = 100 fb$^{-1}$}
\psfrag{NbEvent}[br][t]{Number of events}
\psfrag{Mll}[r][b]{$M_{\ell\ell}$ (GeV)}
\psfig{figure=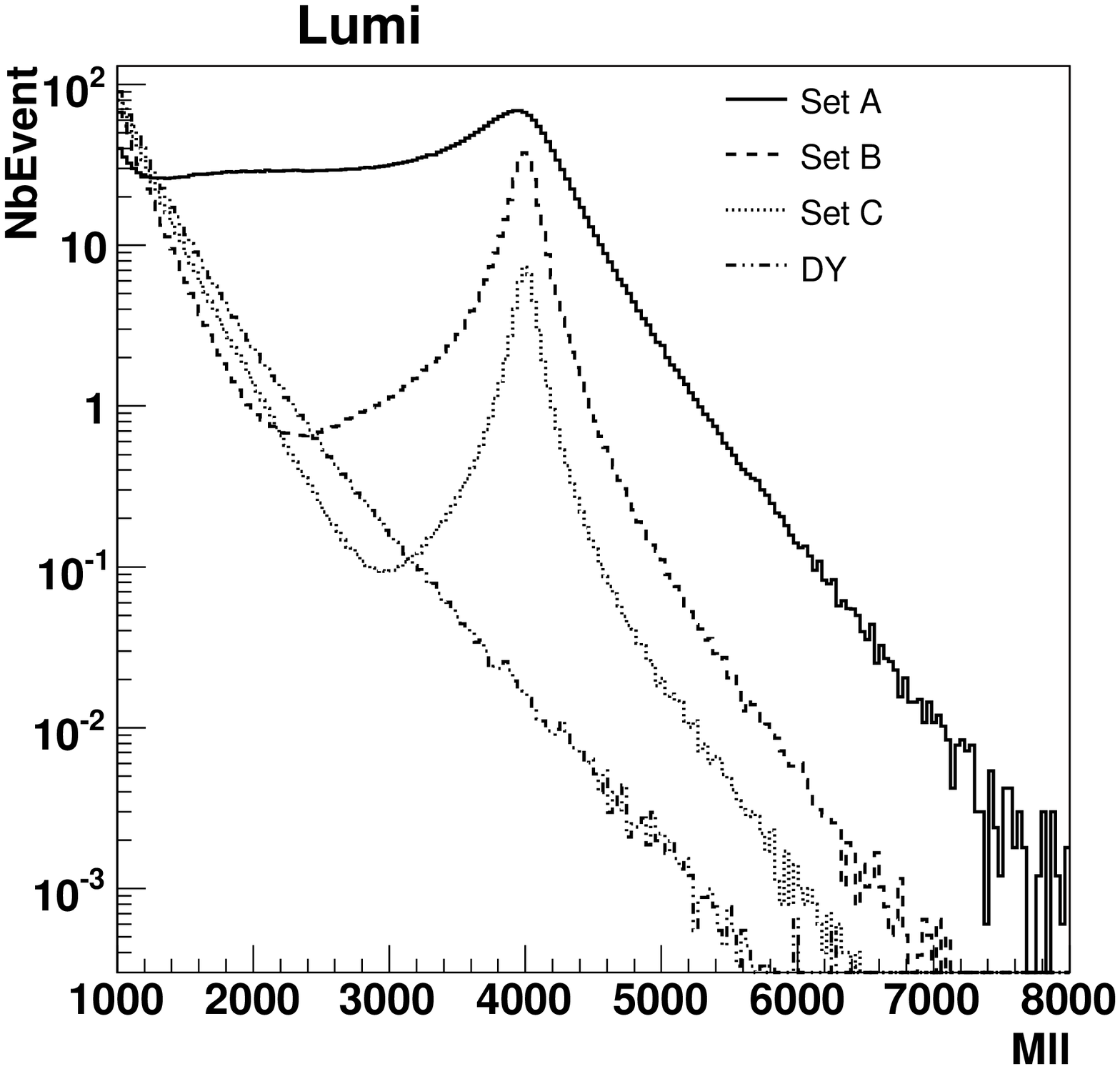,height=7cm}  
\psfrag{Lumi}{}
\psfrag{NbEvent}[br][t]{\tiny N. of events}
\psfrag{Mll}[r][b]{\tiny $M_{\ell\ell}$ (GeV)}
\psfig{figure=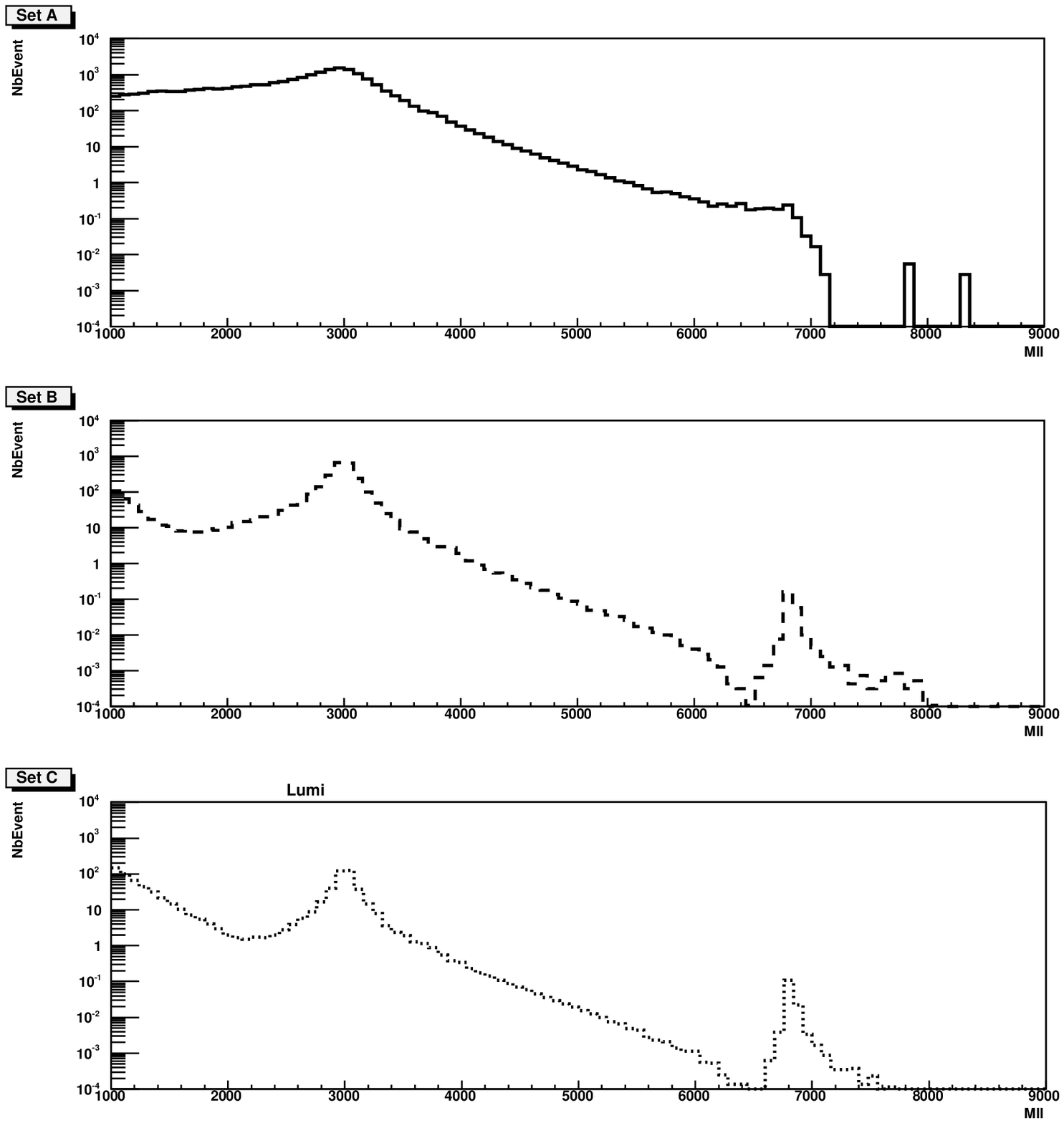,height=6.5cm}  
\end{center} 
\caption{ 
Left: distribution of the generated invariant mass $M_{\ell\ell}$ ($\ell=e$ only or $\mu$ only) for the
{$p p \to \gamma^{(0,1,2)} / Z^{(0,1,2)} \to \ell^+ \ell^-$} process at LHC, with {$M_{KK} = 4$}~TeV  
and parameter sets A (plain line), B (dashed line) or C (dotted line). 
The absolute number of events corresponds to an integrated luminosity of {${\cal L}=100$~fb$^{-1}$}.
The same invariant mass distribution for the pure SM process  
{$p p \to \gamma^{(0)} / Z^{(0)} \to \ell^+ \ell^-$} (dot--dashed line) is also shown.
Right: same distribution with {$M_{KK} = 3$}~TeV.  
\label{distrib}} 
\end{figure} 

Figure~\ref{distrib} (left) shows the generated distribution of the final state di--lepton invariant mass 
$M_{\ell\ell}=\sqrt{(p_{\ell^+}+p_{\ell^-})^2}$ obtained for sets A, B, C with
$M_{KK}=4$ TeV.   
The resonance peak around $M_{\ell\ell}=M_{KK}$ is clearly visible
above a relatively small physical background, the SM Drell--Yan process.
Moreover, the 
$p p \to \gamma^{(0,1,2)} / Z^{(0,1,2)} \to \ell^+ \ell^-$ process yields a large number of events for an  
integrated luminosity ${\cal L}=100$~fb$^{-1}$, which corresponds to one year of LHC running at high luminosity.
Even lower integrated luminosities would lead to a significant number of events.
The difference of KK gauge boson widths between the three parameter sets
originates from the difference in KK gauge couplings. 
It must be noticed also that there is a destructive interference between the SM and
RS contributions which reduces
the number of events, with respect to the pure Drell--Yan process, 
at invariant masses lower than the resonance level.\s

Figure~\ref{distrib} (right) shows the generated distribution of the final state di--lepton invariant mass 
for $M_{KK}=3$ TeV for the three parameter sets separately. 
The second resonance peak, due to the exchange of $\gamma^{(2)}$ and $Z^{(2)}$ excitations,
appears around $M_{\ell\ell}=(5.57/2.45)M_{KK}$.
Its experimental detection would be characteristic of a tower of massive KK states, and would thus represent a strong
indication for the existence of extra dimensions.
Together with a measurement of the $\gamma^{(2)}/Z^{(2)}$ mass, it would constitute a clear signature of the specific RS
model with bulk matter. 
However, the amplitude for $\gamma^{(2)}/Z^{(2)}$ production is highly suppressed by the
decrease of PDFs at large parton energies.\s

\subsection{Detectability}

In order to study the detectability of such events at LHC, the expected
performance of the ATLAS detector~\cite{atlas} has been used.
This performance has been computed using a full simulation of the detector response~\cite{athena}.
The response to the particles out of the tracking acceptance (i.e. with a pseudo-rapidity $|\eta|>2.5$) was not simulated.
The events were then reconstructed in the official ATLAS reconstruction framework~\cite{athena}.\s

We concentrate here on the electron final state, which we have already studied in
detail in the framework of other models~\cite{tev4lhc}.
The muon and tau lepton cases will be commented at the end of this section.
A $\gamma^{(n)}/Z^{(n)}\to e^+e^-$ event selection and reconstruction
is designed and the efficiency of such a selection is 
evaluated as explained in the next subsection.
Finally, the ATLAS discovery reach is computed, as shown in the last subsection.\s 

\subsubsection*{Event selection and selection efficiency}

The same selection as in~\cite{tev4lhc} is used.
First the electron (positron) candidates are reconstructed using the standard ATLAS electron identification:
additionally to criteria on shower shape and energy leakage, one requires to have a good track quality.
The absence of any additional track in a broad cone around the matched track is also required in order to reduce
the QCD and tau backgrounds. 

Only events with at least two electron candidates are selected. 
These two candidates are also required to be isolated in the calorimeter, which means that no more than 40~GeV
have been deposited in the calorimeter in a cone of 
radius $\sqrt{(\Delta\eta)^2+(\Delta\phi)^2}=0.5$
around the electron direction.
Finally, the two electrons are required to be of opposite charge and back to back
in the plane transverse to the beam, the absolute difference of azimuthal angles having to be greater than 2.9 radians.

These criteria are aimed at selecting di--electron events and rejecting possible
background events. After this selection, Drell--Yan events, indistinguishable from $\gamma^{(n)}/Z^{(n)}$ events
are expected to be the only physical background. Some non--physical, reducible background
could come from processes such as $\gamma W$ events in which the photon is misidentified
as an electron and the $W$ decays into an electron. Given their cross section and the rejecting power
of the electron identification, they are assumed to be negligible. 

\begin{figure} 
\begin{center} 
\psfrag{eff}[br]{Efficiency}
\psfrag{mll}[r][b]{$M_{ee}$ (GeV)}
\psfig{figure=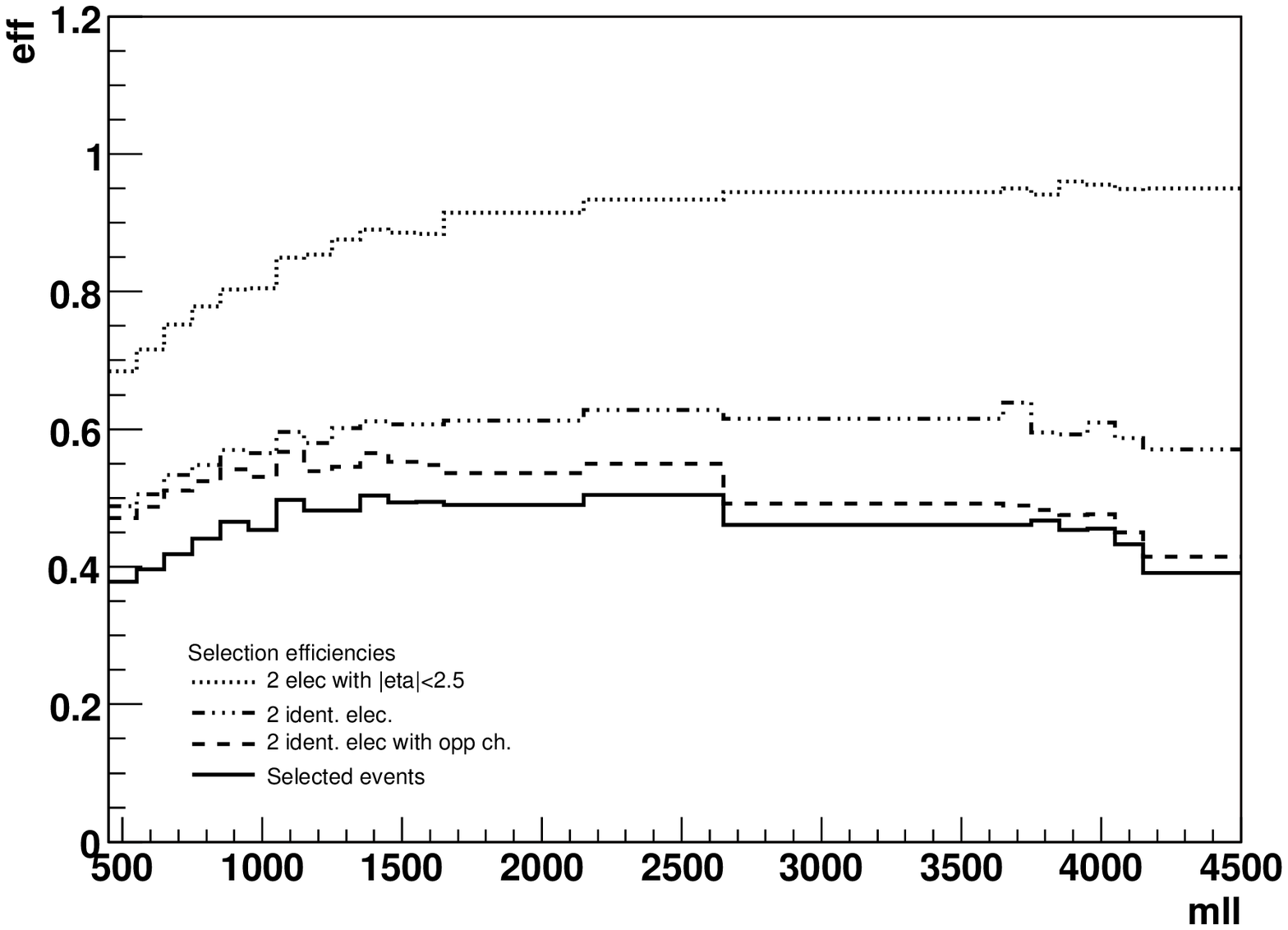,height=5.5cm} 
\psfig{figure=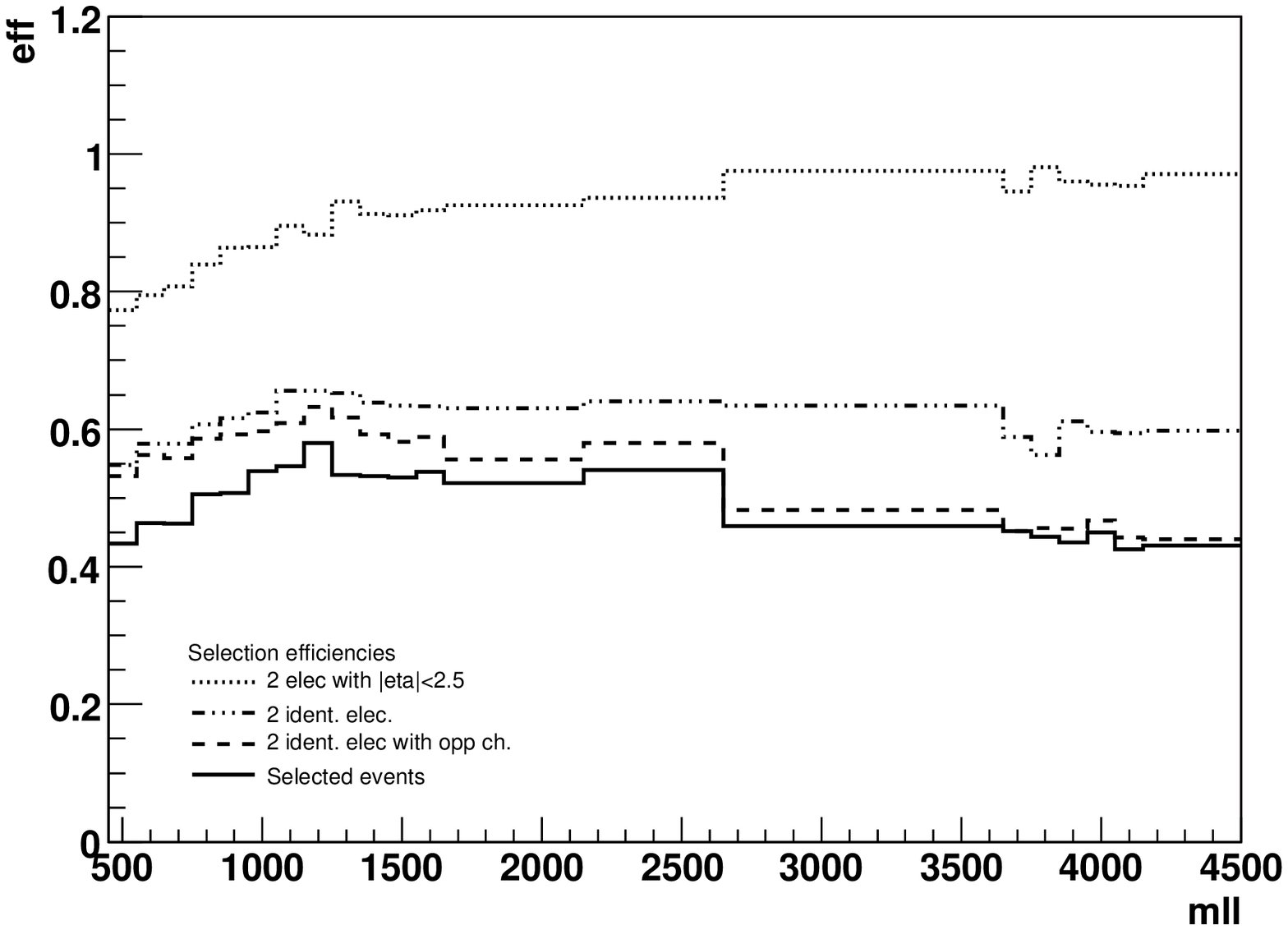,height=5.5cm} 
\end{center} 
\caption{ 
Selection efficiency as a function of $M_{ee}$; 
left: $u\bar u$ events, right: $d\bar d$ events.}
\label{fig:eff}
\end{figure}

The final efficiency of the selection on signal events is shown 
as a function of the di--electron invariant mass on Fig.~\ref{fig:eff}.
Two curves are shown separately for $u\bar u$ and $d\bar d$ events because 
the events arising from $u\bar u$ fusion are slightly more boosted
than those arising from $d\bar d$ fusion (because of their PDFs). 
Provided that one separates these two contributions, it has been shown that 
the selection efficiencies are model independent~\cite{tev4lhc}.
In both cases, the efficiency is relatively flat as a function of the di--electron invariant mass.
No electron was simulated above 4.5~TeV but the performance is expected to remain about the same for higher energies,
even if this implies some initial adjustments.\s 

\subsubsection*{ATLAS discovery reach}

As seen on the invariant mass distribution, the resonance shows a large bump
which can be detected by searching for an excess of events
above the expected spectrum from the SM process. 
One could also exploit the fact that there is a strong
destructive interference at di--lepton invariant masses lower
than the resonance by looking for a {\it deficit} of events. 
For simplicity sake, we restrict here to the search for an excess, but
we note that the sensitivity could possibly be improved by designing
a search for a deficit.

The expected number of signal events ($S$) and of background events ($B$) is 
evaluated\footnote{More precisely, in order to take into account the interference effects,
$S,B$ are defined from the numbers of events $N$ expected within the SM and RS
extension, as follows: $S=N_{SM+RS}-N_{SM}$ and $B=N_{SM}$.}
in the following invariant mass interval: [$M_{thr},\infty$[,
where $M_{thr}=0.6 M_{KK}$ has been optimized in order to integrate the full
signal in the case of set A, which has the largest natural width.\s

In order to compute $S$ and $B$, events have been generated by {\tt PYTHIA}
and efficiency weighted according to
$M_{\ell \ell}$ and to the incoming quark flavour in order to derive an effective production cross section. 
This procedure was also applied to the irreducible background.  
A significance estimator, called $S_{12}$, was finally used in 
order to extract the discovery reach. This estimator is defined by 
$2S_{12} = 2 (\sqrt{S+B} - \sqrt{B})$; this definition has been shown~\cite{SignifSc12}
to be less optimistic than the usual $S/\sqrt{B}$.
The discovery is claimed if the two following conditions are met:
$2S_{12}>5$ and $S>10$.

In order to make a full computation of the discovery reach, it would be necessary
to consider possible systematic effects. These are of various kinds, either experimental
such as the uncertainty on the integrated luminosity, the electron energy scale, etc,
or theoretical, such as higher order corrections to the cross section computation.
This is beyond the scope of this paper, and will be treated elsewhere~\cite{cscNote}.
The results obtained here are thus dominated by the cross section.

\begin{figure}[t]
  \begin{center}
  \psfrag{mZp}[br]{$M_{KK}$ (TeV)}
  \psfrag{Lumi}[r][b]{Integrated luminosity (fb$^{-1}$)}
  \epsfig{file=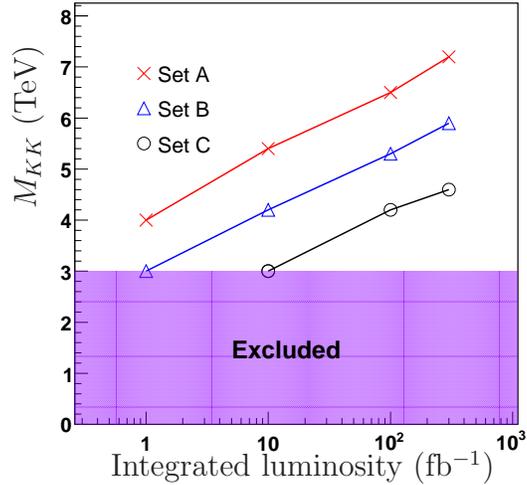,height=7cm}
  \end{center}
  \caption{ATLAS discovery reach in the electron final state in terms of {$M_{KK}$} 
  as a function of the integrated luminosity for parameter sets A, B, C.}
  \label{f_reach}
\end{figure}

The value of the $M_{KK}$ reach is shown as a function of the integrated luminosity on Fig.~\ref{f_reach}.
One can see that the ATLAS discovery potential for the exchange of KK neutral gauge bosons 
is sizable, even for low integrated luminosities. For instance, the medium coupled B set is detectable up to
about 4~TeV with only 10~fb$^{-1}$ of integrated luminosity, which could be reached after a couple of years of running.
The reach extends up to about 5.8~TeV for the same model with 300~fb$^{-1}$.

From the theoretical point of view, the cross sections for electron and muon
productions are almost the same, as explained in Section \ref{Cross}.
Experimentally, a study of the muon detection efficiency based on a fast
simulation has showed that this efficiency should be comparable to the electron one.
Hence, one can 
estimate that including the statistics of the muon final state would be roughly equivalent to 
multiplying the integrated luminosity by a factor of 2, so that the above reaches would be
obtained with twice as less luminosity. 
\newline The rates for electron and tau leptons are also similar. However, the
detection of the tau lepton, which is unstable, is experimentally more difficult than 
the detection of the light stable leptons and would require a specific analysis.
Even if no such specific selection is performed, the leptonic decays of the di--tau final state would contribute 
to the high mass di--lepton spectrum. However, given the branching ratios,
the final significance, and in turn the sensitivity on $M_{KK}$, is not expected to vary significantly.\s

\section{Conclusion}   
\label{conclu}

We have considered several  
configurations of SM fermion localizations, in the RS model, which generate a realistic structure in   
flavour space (reproducing quark/lepton masses and satisfying FCNC bounds for low $M_{KK}$).  
We have noticed that these configurations 
also possess the particularity of 
producing lepton couplings to KK gauge bosons which are larger than in the SM {\it and}  
can remain in agreement with the EW precision data if one assumes a custodial $O(3)$ symmetry. 
Then, based on these different fermion 
configurations, we have shown that the experimental search at LHC for new effects 
in the SM Drell--Yan process coming from exchanges of KK gauge bosons  
would lead to a high sensitivity on $M_{KK}$   
up to $\sim 6$ TeV (depending on the scenario and considered luminosity) 
due to the clean leptonic signature. Such effects would constitute an indication for 
the existence of the O(3) symmetry.\s

\section*{Acknowledgments}  
We thank G. Azuelos and G. Polesello for providing a model of {\tt PYTHIA} user defined process.\\
We thanks members of the ATLAS Collaboration for helpful discussions. 
We have made use of the ATLAS physics analysis framework and tools which are the result of collaboration-wide efforts.\s  
  
\newpage  
  
\section*{Appendix}

We denote set A the following set of $c_i$ values for each SM fermion,   
$$
\begin{array}{cccccc}   
c_{1}^{Q}=0.2\ ;   & c_{2}^{Q}=0.2\ ;  & c_{3}^{Q}=0.2 \ \ \ \ \ \ &  
c_{1}^{L}=-1.5\ ;  & c_{2}^{L}=-1.5\ ;  & c_{3}^{L}=-1.5   
\\   
c_{1}^{d}=0.728\ ;  & c_{2}^{d}=0.740\ ;  & c_{3}^{d}=0.628 \ \ \ \ \ \ &  
c_{1}^{\ell}=0.760\ ;  & c_{2}^{\ell}=0.833\ ;  & c_{3}^{\ell}=0.667   
\\   
c_{1}^{u}=0.62\ ;  & c_{2}^{u}=0.62\ ;\  & c_{3}^{u}=0.35  \ \ \ \ \ \ &  
c_{1}^{\nu}=1.512\ ; & c_{2}^{\nu}=1.513\ ;  & c_{3}^{\nu}=1.468   
\end{array}   
$$
whereas set B is defined by,  
$$   
\begin{array}{cccccc}      
c_{1}^{Q}=0.37\ ;   & c_{2}^{Q}=0.37\ ;   & c_{3}^{Q}=0.37 \ \ \ \ \ \ &  
c_{1}^{L}=0.200\ ;   & c_{2}^{L}=0.200\ ;   & c_{3}^{L}=0.261   
\\   
c_{1}^{d}=0.716\ ;   & c_{2}^{d}=0.728\ ;   & c_{3}^{d}=0.615 \ \ \ \ \ \ &  
c_{1}^{\ell}=0.737\ ;   & c_{2}^{\ell}=0.696\ ;   & c_{3}^{\ell}=0.647   
\\   
c_{1}^{u}=0.607\ ;   & c_{2}^{u}=0.607\ ;  & c_{3}^{u}=0.050  \ \ \ \ \ \ &  
c_{1}^{\nu}=1.496\ ;   & c_{2}^{\nu}=1.503\ ;  & c_{3}^{\nu}=1.463  
\end{array}   
$$   
and set C is given by,
$$   
\begin{array}{cccccc}      
c_{1}^{Q}=0.413\ ;   & c_{2}^{Q}=0.413\ ;  & c_{3}^{Q}=0.413 \ \ \ \ \ \ &  
c_{1}^{L}=0.35\ ;  & c_{2}^{L}=0.35\ ;  & c_{3}^{L}=0.39   
\\   
c_{1}^{d}=0.703\ ;  & c_{2}^{d}=0.721\ ;  & c_{3}^{d}=0.608 \ \ \ \ \ \ &  
c_{1}^{\ell}=0.728\ ;  & c_{2}^{\ell}=0.694\ ;  & c_{3}^{\ell}=0.636   
\\   
c_{1}^{u}=0.60\ ;  & c_{2}^{u}=0.60\ ;\  & c_{3}^{u}=-0.08  \ \ \ \ \ \ &  
c_{1}^{\nu}=1.49\ ; & c_{2}^{\nu}=1.49\ ;  & c_{3}^{\nu}=1.45   
\end{array}   
$$

\end{document}